\documentclass[final,3p,times,twocolumn]{elsarticle}

\usepackage{amssymb}
\usepackage{amsthm}
\usepackage{amsmath}

\usepackage{lineno,hyperref}
\usepackage{graphicx}
\usepackage{epsfig}
\usepackage{booktabs}
\usepackage{footnote}
\usepackage{ctable}
\usepackage{tabularx}
\usepackage{caption,subcaption}
\usepackage{dcolumn}
\usepackage{bm}

\newcommand {\beq} {\begin{eqnarray}}
\newcommand {\eeq} {\end{eqnarray}}
\newcommand {\eeqn} [1] {\label{#1} \end{eqnarray}}

\modulolinenumbers[5]

\journal{Physics Letters B}

\begin{document}

\begin{frontmatter}



\title{Re-examining the transition into the N=20 island of inversion: structure of $^{30}$Mg}


\author[lpl,lpc,usc]{B. Fern\'{a}ndez-Dom\'{i}nguez}\author[lpl]{B. Pietras}\author[surr]{W.N. Catford}\author[lpc]{N.A. Orr}\author[lpl,tud,yor]{M. Petri}\author[lpl]{M. Chartier}\author[lpl,yor]{S. Paschalis}\author[surr]{N. Patterson}\author[surr]{J.S. Thomas}\author[usc]{M. Caama\~{n}o}\author[cns]{T. Otsuka}\author[uam]{A. Poves}\author[cns]{N. Tsunoda}\author[lpc]{N.L. Achouri}\author[lpc]{J-C. Ang\'{e}lique}\author[bham]{N.I. Ashwood}\author[tam]{A. Banu\fnref{pa1}}\author[lpc]{B. Bastin}\author[ifin]{R. Borcea}\author[yor]{J. Brown}\author[lpc]{F. Delaunay}\author[ipno]{S. Franchoo}\author[bham]{M. Freer}\author[ganil]{L. Gaudefroy}\author[tud]{S. Heil}\author[dar]{M. Labiche}\author[lpc]{B. Laurent}\author[dar]{R.C. Lemmon}\author[lbl]{A.O.~Macchiavelli}\author[ifin]{F. Negoita}\author[lpl]{E.S. Paul}\author[ganil]{C. Rodr\'{i}guez-Tajes}\author[ganil]{P. Roussel-Chomaz}\author[ifin]{M. Staniou}\author[yor]{M.J. Taylor\fnref{pa2}}\author[tam,ifin]{L. Trache}\author[surr]{G.L. Wilson}

\address[lpl]{Department of Physics, University of Liverpool, Liverpool, L69 7ZE, UK}
\address[lpc]{LPC-Caen, IN2P3/CNRS, ENSICAEN, UNICAEN et Normandie Universit\'{e}, 14050 Caen Cedex, France}
\address[usc]{Universidade de Santiago de Compostela, 15754 Santiago de Compostela, Spain}
\address[surr]{Department of Physics, University of Surrey, Guildford GU2 5XH, UK}
\address[tud]{Institut f\"ur Kernphysik, Technische Universit\"at Darmstadt, 64289 Darmstadt, Germany}
\address[yor]{Department of Physics, University of York, Heslington, York YO10 5DD, UK}
\address[cns]{CNS, University of Tokyo, 7-3-1 Hongo, Bunkyo-ku, Tokyo, Japan}
\address[uam]{Departamento de Fisica Te\'{o}rica and IFT-UAM/CSIC, Universidad Aut\'{o}noma de Madrid, E-28049 Madrid, Spain}
\address[bham]{School of Physics and Astronomy, University of Birmingham, Birmingham, B15 2TT, UK}
\address[tam]{Cyclotron Institute, Texas A\&M University, College Station, TX-77843, USA}
\address[lbl]{Nuclear Science Division, Lawrence Berkeley National Laboratory, Berkeley, California 94720, USA}
\address[ifin]{IFIN-HH Bucharest-Magurele, RO-077125, Romania}
\address[ipno]{IPN-Orsay, 91406 Orsay, France}
\address[ganil]{GANIL, CEA/DRF--CNRS/IN2P3, BP 55027, 14076 Caen Cedex 05, France}
\address[dar]{Nuclear Structure Group, CCLRC Daresbury Laboratory, Daresbury, Warrington WA4 4AD, UK}
\fntext[pa1]{Present Address: Department of Physics and Astronomy, James Madison University, Harrisonburg, VA 22807, USA}
\fntext[pa2]{Present Address: Division of cancer sciences, University of Manchester, Manchester, M13 9PL, UK}

\begin{abstract}

Intermediate energy single-neutron removal from $^{31}$Mg has been employed to investigate the transition into the N=20 island of inversion.  
Levels up to 5~MeV excitation energy in $^{30}$Mg were populated and spin-parity assignments were inferred from the corresponding longitudinal momentum distributions and $\gamma$-ray decay scheme.  Comparison with eikonal-model calculations also permitted spectroscopic factors 
to be deduced.  Surprisingly, the 0$^{+}_{2}$ level in $^{30}$Mg was found to have a strength much weaker than expected in the conventional
picture of a predominantly $2p - 2h$ intruder configuration having a large overlap with the deformed $^{31}$Mg ground state.  In addition, negative parity levels were identified for the first time in $^{30}$Mg, one of which is located at low excitation energy.
The results are discussed in the light of shell-model calculations employing two newly developed approaches
with markedly different descriptions of the structure of $^{30}$Mg.
It is concluded that the cross-shell effects in the region of the island of inversion at Z=12 are 
considerably more complex than previously thought and that
$np - nh$ configurations play a major role in the structure of $^{30}$Mg.

\end{abstract}


\end{frontmatter}



The ``island of inversion'' (IoI) in which the neutron-rich N$\approx$20 isotopes of Ne, Na and Mg exhibit ground states dominated by cross-shell intruder configurations, has attracted much attention since the first observations \cite{Thibault,Detraz}.  In particular, this region has become the testing ground for our understanding of many of the concepts of shell evolution away from $\beta$-stability and has sparked the development of sophisticated shell-model interactions.  Theoretical approaches first employed mean field \cite{Campi} and, later, shell-model calculations \cite{PovesReta,Warburton,Utsuno1,Otsuka,Utsuno2,Nowacki} to explain the enhanced binding energies and low $2^+$ excitation energies, wherein deformation and a diminished N=20 shell gap \cite{Caurier} result in $fp$-shell intruder configurations dominating the ground state wave functions.

In the case of the Mg isotopes, $^{30,31}$Mg were first suggested to lie outside the IoI, based on their masses \cite{Orr91,Zhou91}.  Subsequent measurements, notably the measurements of the ground state spin-parity ($J^\pi$=1/2$^+$) and magnetic moment \cite{Neyens,Kowalska} have combined with theoretical work (e.g., Refs. \cite{Kimura,Caurier}) to produce a widely accepted picture in which $^{31}$Mg is the lightest magnesium isotope within the IoI.  Its ground state is characterised by a strongly prolate deformed intruder structure with an almost pure neutron $2p-2h$ configuration \cite{Marechal,Marina}. In contrast, $^{30}$Mg is firmly placed outside the IoI
and its structure interpreted as a spherical $0p-0h$ ground state \cite{Terry} coexisting with a neutron $2p-2h$ intruder-dominated deformed 0$^{+}_{2}$ isomeric state at 1.788~MeV \cite{Mach,Schwerd} and with negative parity levels expected to appear, according to shell model calculations, at a relatively high excitation energy ($>$3.5~MeV \cite{Deacon}).

Very recently, calculations employing a new type of interaction -- EEdf1 -- have reproduced many of the 
properties of the neutron-rich isotopes of Ne, Mg and Si \cite{NTsunoda}.  Significantly, the interaction was derived for the $sd + pf$ shells from fundamental principles and explicitly including three-body forces.  Intriguingly, the EEdf1  calculations predict that multiple particle-hole excitations play a much bigger role than suggested by the earlier calculations. For example, in the Mg isotopic chain the admixture of neutron $2p-2h$ and $4p-4h$ configurations increases suddenly at N=18 \cite{NTsunoda}.  Indeed, the ground state structure of $^{30}$Mg is predicted to be very strongly influenced by the intruder $fp$-shell configurations, with $\sim$75\% of the ground state wavefunction being of this nature \cite{NTsunoda}.

In order to test these two very different pictures of the transition into the IoI the structural overlaps between the $^{31}$Mg and $^{30}$Mg states are of critical importance.
To date, however, there are only indirect estimates, based on proton resonant elastic scattering on $^{30}$Mg \cite{IAR}. In the present work, intermediate energy single-neutron removal from $^{31}$Mg is investigated.  In addition to providing a measure of the overlaps between the $^{31}$Mg ground state and the levels populated in $^{30}$Mg, the spins and parities of previously known and newly observed states are deduced.

The experiment was performed at the GANIL facility where a high intensity $^{36}$S primary beam (77.5 MeV/nucleon)
was employed, in conjunction with the SISSI device \cite{SISSI}. A beam analysis spectrometer delivered a secondary beam of $^{31}$Mg (55.1~MeV/nucleon) with a rate of $\sim$55 pps.  The secondary beam bombarded
a carbon target (thickness 171~mg/cm$^{2}$) and the beam-like residues were analysed according to momentum using the SPEG spectrometer \cite{SPEG} and identified in mass and charge using standard $\Delta$E-E-TOF techniques.

The $\gamma$-rays emitted by the beam-like residues were detected using an array of 8 EXOGAM Ge clover detectors \cite{EXOGAM} that were arranged symmetrically in two rings, each of 4 detectors, at polar angles of 45$^{\circ}$ and 135$^{\circ}$ with respect to the beam axis.  The full-energy peak efficiency for the array, after implementing add-back, was measured to be 3.3$\pm$0.1\% at 1.3 MeV and the energy resolution, after Doppler correction, was 2.7\%.  A more complete account of the experimental details may be found in Ref. \cite{ABanu}.

\begin{figure}[ht]
  \includegraphics[width=\columnwidth]{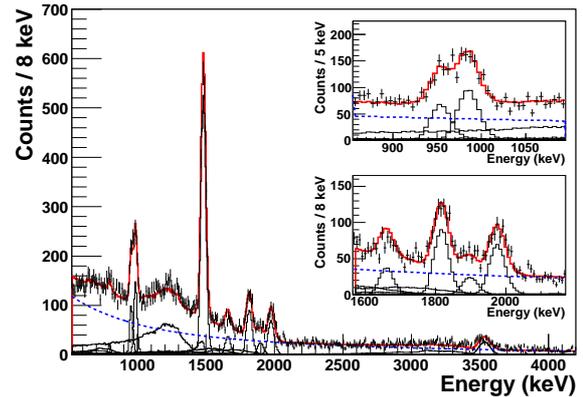}
\caption{(Color online) Doppler corrected and add-back reconstructed $\gamma$-ray energy spectrum (E$_{\gamma}>$ 500~keV) in coincidence with $^{30}$Mg. The overall fit (red line) includes Geant4 generated lineshapes for each transition (black histograms) and an exponential background (blue dashed line). The insets show the details of the regions from 850 to 1100~keV and 1575 to 2175~keV.}
\label{tot_gamma_spectrum}
\end{figure}

\begin{figure}[ht]
  \includegraphics[width=\columnwidth]{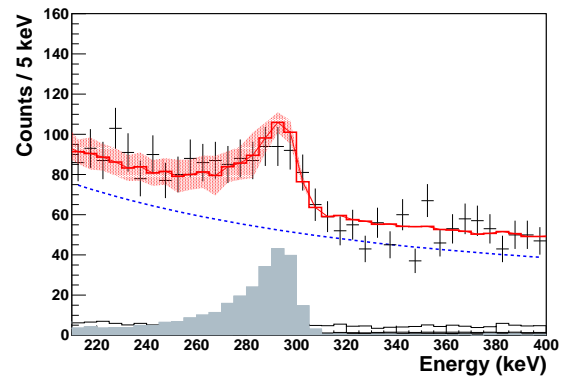}
\caption{(Color online) Spectrum for the forward angle EXOGAM detectors for E$_\gamma<~$500~keV. The grey histogram is the simulated lineshape for the $0^+_2$$\rightarrow$2$^+_1$ decay -- E$_\gamma$=300$\pm$5~keV. The red shading reflects the uncertainty in the half-life -- 3.9$\pm$0.5 ns \cite{Mach}.}
\label{gamma_spectrum_isomer}
\end{figure}

\begin{figure}[ht]
  \includegraphics[width=\columnwidth]{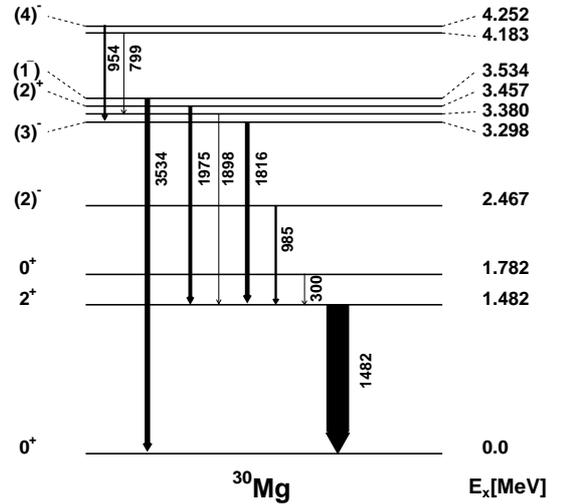}
\caption{Level scheme of $^{30}$Mg obtained in the present work.  The widths of the arrows reflect the absolute intensities of the transitions. The spin-parity assignments above 2~MeV are those deduced here (see text and Table~\ref{excxs}).}
\label{lscheme_gamma}
\end{figure}

The inclusive cross section for single-neutron removal from $^{31}$Mg was determined to be 90$\pm$12 mb where the error arises principally from the uncertainty in the integrated secondary beam intensity. The $\gamma$-ray spectrum, for events observed in coincidence with $^{30}$Mg residues, is shown in Fig.~\ref{tot_gamma_spectrum}, after Doppler and add-back corrections were applied.  Nine known transitions \cite{Bauman,Deacon,Shimoda} were observed. The energies and intensities are listed in Table~\ref{excxs} and the deduced decay scheme shown in Fig.~\ref{lscheme_gamma}.  A further weak, previously unreported transition, which is not in coincidence (within the statistics) with any other $\gamma$-ray line, was identified at 1660(2)~keV.  The $\gamma$-ray energy spectrum was fitted with lineshapes generated for each transition using Geant4 \cite{GEANT4}, plus a smooth continuum background. 

Below 500 keV, no $\gamma$-ray lines were observed other than an asymmetric peak at $\sim$300 keV corresponding to the known 306~keV transition from the isomeric 0$^{+}_{2}$ 1789 keV level to the 2$^{+}_{1}$ state (half-life 3.9(5)~ns \cite{Mach}).
The lineshape for the isomeric decay was simulated (Fig.~\ref{gamma_spectrum_isomer}) using Geant4 and taking into account the half-life and the $^{30}$Mg post-target velocity ($\beta=0.303$).  The analysis employed only the data acquired with the forward four detectors as the corresponding lineshape exhibited particular sensitivity to the lifetime.

The $^{30}$Mg level and $\gamma$-decay scheme in Fig.~\ref{lscheme_gamma} is in
accord with previous studies \cite{Bauman,Deacon,Shimoda}, with the exception
of two previously reported transitions at 990~keV and 1060~keV \cite{Deacon} for which no evidence was found in the present work (insets Fig.~\ref{tot_gamma_spectrum}) or indeed in previous studies \cite{Bauman,Shimoda}.
The $\gamma$-ray intensities were determined by using the lineshapes from the simulations and then correcting the counts in the full energy peak for the efficiency of the Ge array.  The branching ratio for the direct population of each level via the neutron-removal from $^{31}$Mg was obtained by gating on gamma-ray energy and with feeding corrections taken into account.
For the $0^+_2$ isomeric state, the direct feeding was deduced from the gamma-ray decay via the E2 radiative transition since the ratio E0/E2 is very small ($\sim$1.4$\times$10$^{-2}$), given the partial lifetime of the E0 decay ($\tau$($E$0)=396 ns \cite{Schwerd}).
The exclusive cross section for each state is the product of the direct branching ratio and the inclusive cross section. The results are included in Table~\ref{excxs}.

The cross section, $\sigma_{-1n}$, to remove neutrons ($n\ell j$) from a projectile of mass $A$ populating final states $J^{\pi}$ may be expressed theoretically as \cite{Hansen2003},

\begin{equation}
\sigma_{-1n} = \sum_{n\ell j} \left(\frac{A}{A-1}\right)^N C^2S(J^{\pi},n\ell j) \sigma_{\text{sp}}(n\ell j,S_n^{\text{eff}}),
\end{equation}

where $\sigma_{\mathrm{sp}}$ is the single-particle cross section, $[A/(A-1)]^N$ is the center-of-mass correction
($N = 2n + \ell$) \cite{Dieperink1974},
and $S_n^{\text{eff}}$ = $S_n$ + E$_x$ is the effective separation energy
($S_n (^{31}\textrm{Mg}) = 2.310\pm0.005$~MeV \cite{AME2012}) with E$_{x}$ the excitation energy of the state in the A-1 system.

The single-particle cross sections and momentum distributions were computed using the eikonal formalism \cite{codeJAT,Tostevin1,Tostevin2}. The potentials for the neutron-target and core-target interactions were derived using the Jeukenne, Lejeune and Mahaux (JLM) \cite{JLM} nucleon-nucleon effective interaction. The wavefunction of the removed neutron was calculated in a Woods-Saxon potential where the depth was adjusted to reproduce $S_n^{\text{eff}}$.  The Woods-Saxon radius was constrained using Skyrme Hartree-Fock calculations (Sk20 interaction) and the diffuseness set to 0.7~fm.  The $^{12}$C target nucleus was taken to have a Gaussian matter density with a root-mean-square radius of 2.32 fm. 

Compilations of the results of intermediate-energy single-nucleon removal suggest that there is a systematic variation in the ratio of the experimental and theoretical cross sections -- the so-called quenching factor, $R_s$ \cite{Rs_Gade} -- depending on the relative binding energies of the neutron and proton \cite{Rs_update}.  As the exact origins of this quenching remain to be properly elucidated and may well involve a combination of the structure inputs and reaction theory, no attempt is made here to renormalise the experimental spectroscopic factors.  
In addition it may be noted, that the $S_n^{\text{eff}}$ of the levels populated here correspond to $R_s \approx$ 0.85 -- 0.90, and any correction, if valid, would be smaller than the present uncertainties.

\begin{table*}
\setlength\abovecaptionskip{0pt}
\setlength\belowcaptionskip{8pt}
\centering
\caption{\label{excxs} Results for single-neutron removal from $^{31}$Mg. The observed levels (E$_{x}$), transition energies (E$_{\gamma}$), intensities (I$_{\gamma}$), direct fractional population ($b$) and corresponding cross sections ($\sigma_{-1n}$) are listed.  The orbital angular momentum ($\ell$) of the removed neutron, the corresponding chi-squared per degree of freedom ($\chi^{2}/{\nu}$), inferred spin-parity (J$^{\pi}$), theoretical single-particle cross section ($\sigma_{sp}$) and the deduced spectroscopic factor C$^{2}$S$_{exp}$ are also provided.}

\begin{tabular}{cccccccccc}
  \hline\hline
  E$_{x}$ & E$_{\gamma}$ & I$_{\gamma}$ & $b$  & $\sigma_{-1n}$ & $\ell$ & $\chi^{2}/{\nu}$ & J$^{\pi}$ &  $\sigma_{sp}$ & C$^{2}$S$_{exp}$$^{a)}$  \\
   $[$MeV$]$ &  $[$keV$]$     & (\%)        & (\%) &  $[$mb$]$        &      &   &       &     $[$mb$]$      &                   \\
  \hline
  0.0      &   --    &   --      & 28.3(39) $^{b)}$ &  25.5(48)$^{b)}$ &  0 & 2.9 & 0$^{+}$     & 56.4   & [0.42(8)]$^{b)}$  \\ 
  1.482(2) & 1482(2) & 57.0(28)  & 12.2(33) $^{b)}$ &  11.0(33)$^{b)}$ &  2 & 3.0 & 2$^{+}$     & 23.4  & [0.44(13)]$^{b)}$  \\
  1.782(5) &  300(5) & 8.6(13)   & 8.6(13)          &   7.7(15)        &  0 & 0.5 & 0$^{+}$     & 36.9  & 0.20(4) \\
  2.467(3) &  985(2) & 8.4(5)    &  8.4(5)          &   7.6(11)        &  1 & 1.0 & (2)$^{-}$   & 32.8  & 0.21(3) \\
           &         &           &                  &                  & (2)&(1.6)&             &       &         \\
  3.298(3) & 1816(2) & 13.0(8)   &  7.6(8)          &   6.8(11)        &  3 & 1.9 & (3)$^{-}$   & 17.6  & 0.35(6) \\
           &         &           &                  &                  & (2)&(2.8)&             &       &         \\
  3.380(3) & 1898(2) & 4.2(4)    &  2.8(4)          &   2.5(5)         & -- $^{c)}$ & --  & --  &  --   & --    \\
  3.457(3) & 1975(2) & 10.6(6)   & 10.6(6)          &   9.5(13)        &  2 & 0.8 & (2)$^{+}$   & 18.7  & 0.48(7) \\
           &         &           &                  &                  & (1)&(2.2)&             &       &        \\
  3.534(6) & 3534(6) & 12.3(9)   & 12.3(9)          &   11.1(16)       &  1 & 1.9 &(1$^{-}$)    & 28.9  & 0.35(5)  \\
           &         &           &                  &                  & (2)&(2.0)&             &       &         \\
  4.183(3) &  799(2) &  1.4(1)   &  1.4(1)          &   1.3(2)         & -- $^{c)}$ & --        & --    & --   \\
  4.252(3) &  954(2) &  5.4(3)   &  5.4(3)          &   4.9(7)         &  3 & 1.4 &(4)$^{-}$    & 16.6  & 0.27(4)  \\
           &         &           &                  &                  & (2)&(1.6)&             &       &         \\
 \hline
 --        & 1660(2) & 2.4(2)    &  2.4(2)          &   2.2(3)         &  -- $^{c)}$ & -- & -- & -- & -- \\ \hline\hline
\end{tabular}
\end{table*}
\begin{table*}
  \vspace{-0.35cm}
\begin{tabular}{l}
$^{a)}$ After accounting for the centre-of-mass correction.  A systematic uncertainty of $\sim$10\% related to the reaction\\
modelling is not included in the quoted error \cite{Rs_update,Sauvan04}.\\
$^{b)}$ Upper limits from observed yields (see text). \\
$^{c)}$ The corresponding momentum distributions could not be extracted.\\
\end{tabular}
\end{table*}

\begin{figure}[ht]
\includegraphics[width=\columnwidth]{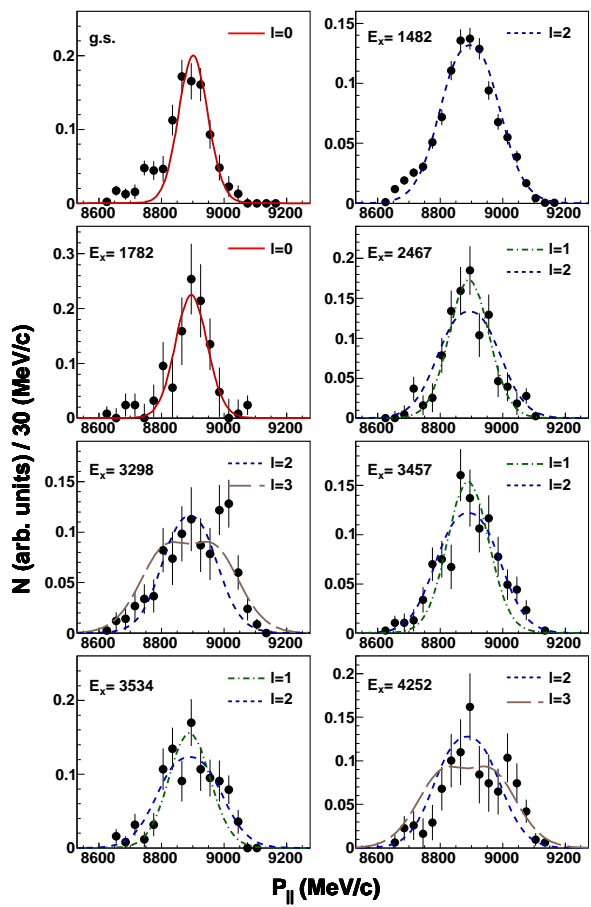}
\caption{(Color online) Exclusive momentum distributions derived for states in $^{30}$Mg, labelled by the excitation energy in keV, compared to eikonal-model predictions; fits were for momenta above 8750 MeV/c (see text).}
\label{pdist}
\end{figure}

The exclusive longitudinal momentum distributions extracted for the ground and excited states are presented in Fig.~\ref{pdist}.  The ground state distribution was obtained from the inclusive results by subtracting, with appropriate weighting, the momentum distributions for the observed excited states.  The theoretical momentum distributions derived from the eikonal model calculations \cite{codeJAT} were folded with the experimental resolution (FWHM=75~MeV/c) and are compared to the experimental results in Fig. \ref{pdist} where the overall normalisation has been adjusted to provide the best possible description\footnote{We believe that this is preferable to the commonly adopted procedure whereby the theoretical lineshapes are normalised to the peak of the experimental distribution.}. (Table~\ref{excxs}).  In order to avoid any bias from dissipative processes in the reaction \cite{Sauvan04} which are not incorporated in the eikonal modelling, the theoretical lineshapes were compared to the data only for momenta greater than 8750~MeV/c, since the inclusive momentum distribution, as well as that for the ground and 2$^+_1$ states, show evidence of tails at momenta below this value.  In the cases of the levels with unknown spin-parities (ie., above the 0$^+_2$ state), the lineshapes for the two $\ell$ values that come closest to reproducing the data are shown.

The most likely $\ell$ values for the removed neutron were deduced for all except two levels (3.534 and 4.252~MeV) according to the smallest $\chi^{2}/{\nu}$ values (Table~\ref{excxs}). The absolute values of $\chi^{2}/{\nu}$ reflect, in addition to statistical variations, contributions from any imperfections in the $\gamma$-ray gating and background substraction, and uncertainties in the theoretical lineshapes\footnote{These may arise from effects not incorporated in the model employed here \cite{deformed_mtm_dist,ES_JAT_deformed}, the choice of the model or formalism (see, for example, refs.~\cite{Sudden,Transfer-to-cont}) and uncertainties in the parameters of the model itself.} as evidenced by the comparatively poor $\chi^{2}/{\nu}$ for the 2$^+_1$ level.

\begin{figure*}[ht]
\includegraphics[width=\textwidth]{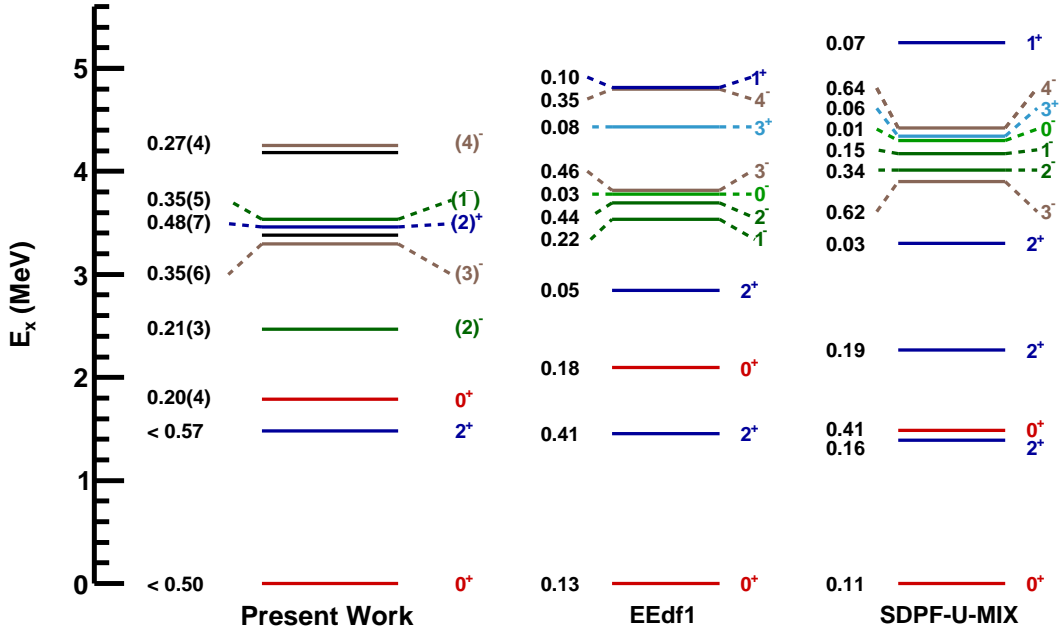}
\caption{(Color online) Level scheme for $^{30}$Mg obtained in the present work compared to shell model calculations using the EEdf1 \cite{NTsunoda} and SDPF-U-MIX \cite{Caurier} interactions. Spectroscopic factors (see text and Table~\ref{excxs}) and proposed spin-parity assignments are shown on the left and right side of the energy levels, respectively. The energies of the experimentally observed states are listed in Table 1 and Figure \ref{lscheme_gamma}. The orbital angular momentum ($\ell$) for the removed neutron is also indicated: $\ell$=0 (red), 1 (green), 2 (blue), 3 (brown), undetermined (black) - see Table \ref{excxs}.}
\label{lscheme_all}
\end{figure*}

Removing a neutron from the 1$s_{1/2}$ orbital in $^{31}$Mg leads to 0$^+$ (and potentially 1$^+$) states in $^{30}$Mg.  In the case of the $0^{+}$ ground state, the deduced spectroscopic factor of 0.42$\pm$0.08 is much larger than the indirect estimate of Imai {\it et al.} \cite{IAR}, namely C$^2$S=0.07$\pm$0.03(stat)$\pm$0.07(sys),  and both of the shell model predictions discussed below (C$^2$S=0.11--0.13).  Given that the cross section to the ground state is derived assuming
that {\it all} of the yield to bound excited levels has been identified (S$_n$($^{30}$Mg)=6.35$\pm$0.01~MeV \cite{AME2012}), both the cross section and the associated spectroscopic factor (Table~\ref{excxs}) should be considered as upper limits.  Indeed, high energy (and unobserved) $\gamma$-rays, would arise from any 1$^+$ levels populated by 1$s_{1/2}$ and/or 0$d_{3/2}$
neutron removal; these are predicted to lie near 5~MeV (Fig.~\ref{lscheme_all}) and candidates are suggested in $\beta$-decay measurements \cite{Shimoda}.

In the case of the 0$^{+}_{2}$ level, the spectroscopic factor of 0.20$\pm$0.04 deduced here is, in terms of the conventional picture of an intruder-dominated $2p-2h$ configuration, surprisingly low.

Removing a neutron from the 0$d_{3/2}$ orbital in $^{31}$Mg can populate 1$^{+}$ or 2$^{+}$ states in $^{30}$Mg.
The 2$^{+}_{1}$ state is populated strongly here, however, the associated spectroscopic factor (0.44$\pm$13) must be interpreted with caution as the deformed character of $^{30}$Mg \cite{Christe} will permit dynamical excitations (and de-excitations) to occur during the neutron removal.  
CCDC-type calculations suggest that such effects
will result in a net increase in the yield to the 2$^{+}_{1}$ state and the spectroscopic factor deduced here should, thus,  
be considered an upper limit~\cite{Summers2006}.

The next highest state characterised by $\ell$=2 neutron removal is found at 3.457~MeV. A fusion-evaporation study \cite{Deacon} suggested a 4$^{+}$ assignment which would, for a single-step process, require neutron removal from the $1g_{9/2}$ orbital.  This is incompatible with any reasonable structure for $^{31}$Mg and with the observed momentum distribution.  Given that any 1$^+$ levels are expected (Fig.~\ref{lscheme_all}) at high energy, a 2$^{+}_{2}$ assignment is made here, in line with the $^{30}$Na $\beta$-decay study \cite{Shimoda}.  The large spectroscopic factor of 0.48$\pm$0.07 suggests an intruder dominated structure.

Negative parity states will be populated via removal of an $fp$-shell neutron from $^{31}$Mg.  Significantly, the level
at 2.467~MeV has a momentum distribution characteristic of $\ell$=1 (1$p_{3/2}$) removal, for which spin-parities of 1$^-$ and 2$^{-}$ are possible.
Given that the $\gamma$-decay proceeds to the 2$^+_1$ level, a 2$^{-}$ assignment is clearly favoured\footnote{Whereas this contradicts Ref.~\cite{Mach}, the feeding in $^{30}$Na $\beta$-decay appears to be forbidden \cite{Deacon,Shimoda}, thus implying negative parity.} since a 1$^{-}$ assignment would favor direct E1 decay to the ground state.  The presence of a negative parity state at such a low energy is surprising in view of the shell model predictions (Fig~\ref{lscheme_all}) and in comparison with the corresponding levels in $^{26,28}$Mg (E$_x$=6.19 and 5.17~MeV).

The momentum distribution for the 3.534~MeV level is compatible with $\ell=2$ or $1$, and thus with spin-parity assignments
of (1, 2)$^{+}$ or (1, 2)$^{-}$ for 0$d_{3/2}$ or 1$p_{3/2}$ neutron removal respectively.
Given, however, that the $\gamma$-decay proceeds only via direct decay to the ground state a spin of 1 is clearly favoured. Furthermore an assignment of $1^{-}$ is strongly suggested since it would be the partner of the nearby 2$^{-}$ state produced by 1$p_{3/2}$ removal.

The levels at 3.298 and 4.252~MeV both exhibit momentum distributions consistent with $\ell=3$ neutron removal, although in the case of the later $\ell=2$ removal is also possible.  The $\ell=3$ (0$f_{7/2}$) removal suggests spin-parities of 3$^{-}$ or 4$^{-}$.  
Significantly, the lower of the two states decays to the $2_1^+$ state with the associated transition being E1 or E3
for the $3^-$ or $4^-$ assignments respectively.  The latter transition would, however, very probably be isomeric with a lifetime approaching
1~$\mu$s, indicating that the lower level is 3$^{-}$. For the higher lying level $\ell$=2 neutron removal suggests spin-parities of (1, 2, 3)$^{+}$.  Such assignments would be expected to be characteristed by M1 decays \cite{Endt_gamma} to the low lying 0$^{+}$ and 2$^{+}$ states rather than the observed decay to the 3.298~MeV 3$^-$ level.  As such it may be reasonably concluded that the 4.252~MeV level is populated as the 4$^-$ spin-coupling partner of the 3$^-$ state.

Turning now to the interpretation of the results presented here, Fig.~\ref{lscheme_all} provides a comparison with shell model
calculations\footnote{The EEdf1 calculations were performed in the $sdpf$ model space for both protons and neutrons, while for the SDPF-U-MIX the protons were confined to the $sd$ shell.} using the recently developed EEdf1 \cite{NTsunoda} and SDPF-U-MIX \cite{Caurier} interactions.  The former was derived from
chiral effective field theory nucleon-nucleon interactions, whilst the latter is an extension of the SDPF-U interaction \cite{Nowacki} which allows for the mixing of different $np-nh$ configurations.
It should be noted that while the $^{31}$Mg ground state is essentially of intruder character in both calculations, the details differ markedly: 90\% of the SDPF-U-MIX wave function is $2p-2h$, whilst, in the case of EEdf1, 66\% is $2p-2h$ and 29\% $4p-4h$.

\begin{figure}[ht]
\includegraphics[width=\columnwidth]{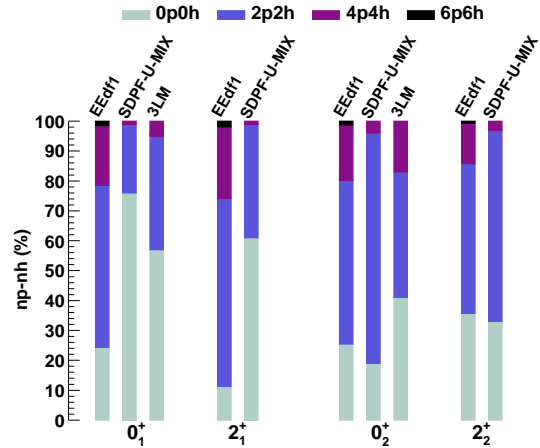}
\caption{(Color online) Decomposition of the wavefunctions in $^{30}$Mg for the 0$^{+}_{1,2}$ and 2$^{+}_{1,2}$ states as derived from shell-model calculations employing the EEdf1 and SDPF-U-MIX interactions and, for the 0$^{+}_{1,2}$ levels, by a three-level mixing model (3LM) -- see text.}
\label{wf_comp}
\end{figure}

The difference between the two shell model calculations is most apparent in the character of the $^{30}$Mg states: the EEdf1 predicts the 0$^{+}_{1}$, 0$^{+}_{2}$, 2$^+_1$ and 2$^+_2$ levels to be overwhelmingly
dominated ($\gtrsim$70\%) by intruder configurations while the SDPF-U-MIX suggests a more conventional situation with only the
0$^{+}_{2}$ and 2$^+_2$ states having intruder character (Figure \ref{wf_comp}).  In contrast, for the negative parity $fp$ shell states, the two calculations agree (Fig. \ref{lscheme_all}) that they should lie at relatively high excitation energy ($\gtrsim$3.5~MeV).

Focussing on the new results, the spectroscopic factors measure the structural overlap of the IoI nucleus $^{31}$Mg with key low-lying levels in $^{30}$Mg. As shown in Fig. \ref{lscheme_all}, the spectroscopic factor deduced here for the 0$^+_2$ state is in very good agreement with the EEdf1 based prediction and at clear variance with that of the SDPF-U-MIX (being a factor $\sim$2 weaker), suggesting the increased importance of $4p - 4h$ configurations. 

In order to explore further the influence of $4p-4h$ configurations, calculations have been made using 
a three level mixing model (3LM) \cite{Augusto}. In this approach, the starting point comprised the unmixed $np-nh$ ($n$=0, 2 and 4) 0$^+$ levels derived from the SDPF-U-MIX interaction. The mixing was then varied until the excitation energy of the 0$^+_2$ state equalled the experimental value.  As may be seen in Figure~\ref{wf_comp}, the 0$^+_2$  level is strongly mixed with a significant $4p-4h$ component (comparable to the EEdf1 prediction).
The spectroscopic factors derived from the overlap of the SDPF-U-MIX ground state for $^{31}$Mg with the $^{30}$Mg 0$^{+}_{2}$ state from the 3LM (C$^{2}$S=0.26) is in better agreement with the experiment than the SDPF-U-MIX prediction. In the case of the ground state, the strength (C$^2$S=0.22) is twice that of the SDPF-U-MIX prediction (Figure~\ref{lscheme_all}).  For completeness, it may be noted that the 0$^{+}_{3}$ level in the 3LM (C$^2$S=0.15) is predicted to lie at around 3.8~MeV.

Turning to the negative parity states, the lowest in energy is the 2$^-$, which is observed to lie well below those predicted by both the EEdf1 and SDPF-U-MIX interactions, with a strength significantly weaker than either of the calculations.  The proposed 1$^-$ spin-coupling partner of the 2$^-$ is found to lie some 1~MeV higher in excitation energy with a considerable strength.  The shell model calculations are reasonable in terms of the excitation energy but underestimate the strength by a factor $\sim$2.  Finally, the 3$^-$ and 4$^-$ levels are both reasonably well reproduced in terms of energy by the two calculations.  However, while the strengths of each are in very good agreement with the EEdf1 predictions, they are considerably over-predicted by the SDPF-U-MIX calculations.  Interestingly, the N=20 shell gap incorporated in the SDPF-U-MIX interaction is around 5.5~MeV, while that of EEdf1 is only 2.8~MeV.

\begin{figure}[ht]
\includegraphics[width=\columnwidth]{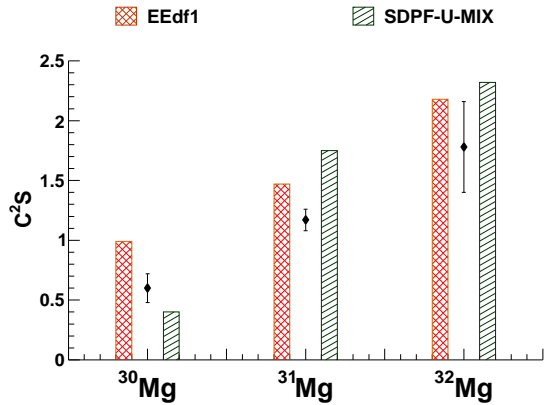}
\caption{(Color online) Comparison of the experimentally deduced summed strength for neutron $fp$-shell levels populated in one-neutron removal from $^{30-32}$Mg  with shell-model predictions using the EEdf1 and the SDPF-U-MIX interactions. The experimental strengths (points with error bars) for $^{30,32}$Mg are taken from Ref.~\cite{Terry}.}
\label{fp_evol}
\end{figure}

Finally, it is instructive to map the evolution of the strength of the $fp$-shell intruder states across the boundary of the IoI.  This is shown in Figure~\ref{fp_evol}, where the summed strength of the negative parity levels observed in the A-1 nuclei populated in single-neutron removal from $^{30-32}$Mg are compared to the shell-model calculations using the EEdf1 and SDPF-U-MIX interactions.
This comparison highlights the enhanced role played by the cross-shell excitations in the EEdf1 calculation, which displays a smooth evolution of structure with neutron number, whereas the SDPF-U-MIX model shows a clear transition at $^{31}$Mg.  The integrated experimental intruder strengths do not permit a definitive choice to be made between the two descriptions of $^{30}$Mg  but do tend to support the smooth evolution predicted by the EEdf1 model.

In conclusion, the structure of $^{30}$Mg has been investigated using single-neutron removal from $^{31}$Mg.  The results, most notably the relatively weak spectroscopic strength for the 0$^+_2$ state and the identification of a low lying negative
parity (2$^-$) level, are at odds with the conventional picture of the transition into the IoI.  Comparisons are made with the results of shell model calculations employing two recently developed interactions with very different descriptions of the underlying structure of $^{30,31}$Mg. These suggest that the low lying levels in $^{30}$Mg are dominated by $np-nh$ configurations, including significant $4p-4h$ contributions.  
As such the transition into the IoI at Z=12 appears to be considerably more complex and less well defined than previously thought.  Ideally, improved measurements should be made (including high-energy $\gamma$-ray detection) so as to clarify the direct population of the ground and 2$^+_1$ states. In addition, an investigation of the d($^{30}$Mg,p) neutron transfer reaction would be valuable.

{\it Acknowledgements}

The authors acknowledge the support provided by the technical staff of LPC and GANIL, including that of their late colleagues and friends J-M.Gautier and J-F. Libin. Thanks are also due to J. Pereira-Conca for fruitful discussions. B.F.D. and M.C.F acknowledge financial support from the Ram\'on y Cajal programme RYC-2010-06484 and RYC-2012-11585 and from the Spanish MINECO grant No. FPA2015-71690-P. W.N. Catford acknowledges financial support from the STFC grant number ST/L005743/1. This work is partly supported by MINECO (Spain) grant FPA2014-57196 and Programme ”Centros de Excelencia Severo Ochoa” SEV- 2012-0249. The participants from the Universities of Birmingham, Liverpool, Surrey and CCLRC Daresbury, GSI and IFIN-HH laboratories also acknowledge partial support from the European Community within the FP6 contract EURONS RII3-CT-2004-06065.

\label{sup}






\end{document}